# Experiments with Compact Antenna Arrays for MIMO Radio Communications

David W. Browne[†], Majid Manteghi[††], Michael P. Fitz[†], Yahya Rahmat-Samii[††]

[†]UnWiReD Laboratory, [††]Antenna Research and Measurements Laboratory

Department of Electrical Engineering

University of California at Los Angeles

Los Angeles, CA 90095-1594

emails: decibel@ee.ucla.edu; majid@ee.ucla.edu; fitz@ee.ucla.edu; rahmat@ee.ucla.edu

*Abstract* – The problem addressed in this study is how to design and test compact antenna arrays for portable Mulitple-Input Multiple-Output (MIMO) transceivers. Mutual coupling in an antenna array affects signal correlation and array radiation efficiency - both of which have dramatic consequences for MIMO channel capacity. Mutual coupling becomes more pronounced as array aperture shrinks and is therefore a critical issue in compact array design. Two novel compact arrays are designed and fabricated for use in MIMO enabled mobile devices. These arrays are extremely compact yet demonstrate acceptable mutual coupling and radiation efficiency because of the MIMO-specific criteria used during their design. An experimental methodology is presented for fair and meaningful characterization of MIMO arrays by field trial. This methodology addresses the issue of capacity normalization and quantifies how well an antenna array's radiation pattern interfaces with multipath propagation. Results are presented from an extensive measurement campaign in which a true MIMO transceiver testbed is outfitted with compact arrays and dipole arrays of various sizes. A comprehensive and fair comparison is made between the compact arrays and dipole arrays in a variety of indoor propagation scenarios. Design recommendations for compact MIMO arrays are given.

## I. INTRODUCTION

Mulitple-Input Multiple-Output (MIMO) radio communications between transceivers with antenna arrays allows information throughput and reliability far greater than what is possible with Single Input Single Output (SISO) communications using single antennas [1], [2]. In the MIMO paradigm, a transmit array with $L_T$ antennas communicating to a receive array with $L_R$ antennas through an i.i.d. Gaussian MIMO channel can achieve an average spectral efficiency that grows linearly as $\mathcal{O}(\min(L_T, L_R))$. This increased performance is possible because multipath propagation between antenna arrays can support independent parallel channels for communicating information. Theoretical models and measurement results show that correlation between the $L_T L_R$ MIMO sub-channels causes a loss of channel capacity relative to the i.i.d. Gaussian MIMO channel [3], [4]. Mechanisms that cause MIMO sub-channel correlation include poor diversity in the propagation multipath and mutual coupling of

This work was supported in part by a United States National Science Foundation research grant (ANI-0305302), by a grant from Marvell Semiconductor through the UC Discovery Program, and by a hardware donation from LEGO Denmark.

signals between elements of the antenna arrays [3], [5]. Both propagation-induced correlation and coupling-induced correlation are dependent on array geometry and become more pronounced as the array's inter-element spacing shrinks [6], [7]. Theoretical work has shown that mutual coupling has a significant effect on MIMO channel capacity [8]-[11]. Whether or not coupling-induced correlation degrades MIMO channel capacity depends on the spatial characteristics of the associated multipath propagation [8]. Sub-channel correlation is a critical issue in portable MIMO enabled devices where the antenna array must be shrunk into a volume whose dimensions are typically measured in fractions of the resonant frequency's wavelength. Another problem resulting from mutual coupling in compact arrays is the subsequent degradation of the array's radiation efficiency [12]. Narrowband methods for combating coupling-induced correlation and impedance mismatch using conjugate matching networks and load matching networks have been developed [13], [14]. Though mathematically tractable, a conjugate matching network that annihilates coupling-induced correlation in an antenna array may be difficult to realize as a compact physical circuit. Still more difficult is the problem of designing and realizing matching networks for multiband and wideband MIMO transceivers.

This work makes a study of compact MIMO arrays with two general contributions: (i) the realization of compact arrays that are designed to preserve MIMO channel capacity without the need for matching networks and (ii) the characterization of compact array performance by field measurements using MIMO radios. A summary of select prior studies relevant to this work puts the current state of progress in perspective.

Of the many studies reporting compact array designs, a fraction report measurements from fabricated arrays and fewer yet evaluate array performance in field trials. Compact diversity arrays using a Printed-Inverted-F Antenna (PIFA) have been reported [15]-[19]. The first four studies present array designs that demonstrate very good diversity performance in measurement and simulation. The arrays in [17] and [18] are similar to the 2-element PIFA array used in this work and the array in [19] but are twice as large (with respect to the resonance wavelength). Compact arrays of three and four antenna have been reported for diversity communications in [20] and [21]. With the exception of [19], none of these studies report on the channel capacity achieved by the arrays in field measurements using MIMO radios.

Channel capacity measurements using compact arrays with inter-element spacings $< \lambda/2$ have been reported [22]-[24]. However, the radios used to make these measurements used time-multiplexed switching between single-port radios and multi-port arrays. Uncorrelated phase noise across antennas in time-multiplexed measurements has been shown to result in optimistic channel capacity results [25]. Furthermore, to capture the effect that mutual coupling has in a wideband MIMO system requires excitation of the entire spectral band on all transmit elements simultaneously and signal acquisition of the entire spectral band from all receiver elements simultaneously. Such a signal excitation and acquisition scheme is true to the actual signaling architecture used in real MIMO transceivers and is termed "true-MIMO". MIMO channel capacity has been characterized as a function of antenna array topology using true-MIMO testbeds [4], [26]-[28]. However, these studies only considered narrowband channels or large arrays with inter-element spacings $\geq \lambda/2$. Only one study is known to have fabricated a wideband compact array with spacing $< \lambda/2$ and measured the channel capacity achieved with it using a true-MIMO testbed [19]. However, the study was limited to a $2 \times 2$ MIMO configuration where the spacing in the other array was sufficiently large to cause very little observable impact on channel capacity.

Several open problems relating to compact array design and testing are addressed in what follows. A novel PIFA design is used as the basic component for building compact arrays because of its low profile and robust performance in the presence of other nearby PIFAs. Two compact arrays are designed and fabricated for use in MIMO enabled mobile devices. During the design process, these arrays are characterized by a metric that measures the effect of mutual coupling on array efficiency under MIMO signaling conditions. The resulting arrays are extremely compact yet demonstrate acceptable mutual coupling and radiation efficiency. An experimental methodology is formulated to allow a fair and meaningful characterization of MIMO arrays by field trial. This methodology addresses the issue of capacity normalization in a way that does not mask the effects of mutual coupling but still allows a fair comparison of antenna array performance from measurements taken in different propagation scenarios. A statistical method is presented for quantifying how well an antenna array's radiation pattern interfaces with multipath propagation by characterizing the effect that an array has on the MIMO channel's underlying eigen-structure. This experimental methodology is implemented in an extensive measurement campaign in which a true-MIMO transceiver testbed is outfitted with compact PIFA arrays and dipole arrays of various size. The resulting measurements quantify the extent to which mutual coupling affects channel capacity. The effect that antenna directionality has on the MIMO channel's eigen-structure is clearly observed in the measurements. The results also present a comprehensive and fair comparison of the channel capacity achieved by the compact PIFA arrays and dipole arrays in a variety of indoor propagation scenarios.

The remainder of the paper is organized as follows. Section II addresses the design and characterization of compact antenna arrays as they relate specifically to MIMO communications. Section III addresses the mathematical framework and experimental methodology used in measuring compact MIMO array performance in field trials. Results from a measurement campaign are presented in Section IV. A summary of the major results and contributions is given in Section V.

## II. COMPACT ARRAY DESIGN

The first general contribution of this work is the realization of compact antenna arrays that are designed to preserve MIMO channel capacity without the need for matching networks. In this section we first present aspects of the MIMO propagation channel as they relate to MIMO antenna array design. We then present a novel low profile antenna design for wideband operation at three resonant frequencies. Compact array designs consisting of two and four elements are then realized from this novel antenna. These arrays are well suited for mobile MIMO terminals because of their low mutual coupling, good radiation efficiency, and compactness. We use a single quantitative measure to characterize both coupling and efficiency in a MIMO array under conditions reflecting those of actual MIMO signaling. Finally, simple dipole arrays are briefly discussed because of their use as reference arrays in this study.

### A. MIMO Channels and MIMO Arrays

A stochastic system model for correlated MIMO channels has been proposed in which the correlation between antenna arrays is characterized jointly [29]. This *joint-correlation channel model* more accurately predicts measured MIMO channel capacity than models which characterize correlation at the transmit and receive arrays independently [30]. The fact that MIMO channel capacity is a function of the joint correlation between transmit and receive arrays suggests that MIMO antenna arrays should not be designed within the diversity-array paradigm in which correlation is mitigated independently in each array by fixed orthogonal division of angle-space or polarization. Nor should MIMO antenna arrays be designed within the smart-antenna paradigm where each array matches its antenna radiation pattern to the multipath it perceives alone. Indeed, such approaches to array design can only hope to achieve the capacity available from the strongest MIMO eigen-channel (though it be greater than SISO channel capacity). Instead, a pair of arrays designed to preserve the available MIMO channel capacity should permit all possible eigen-patterns to be formed jointly between the array-pair and simultaneously for all $\min(L_T, L_R)$ eigen-channels as called for by the channel's Singular Value Decomposition (SVD) so that the full channel capacity of the MIMO channel can be utilized [31].

A subtle but important characteristic of the correlated MIMO channel is that coupling-induced correlation in only one array affects signal correlation in the transmit and receive arrays jointly. This point has been ignored in MIMO mutual coupling studies using *marginal-correlation channel models* which characterize correlation at the transmit and receive arrays independently or *near-marginal-correlation channel models* in which only one array is immersed in rich multipath. It has been shown that mutual coupling in a single antenna array can decorrelate received signals by influencing the array's angle diversity and that this decorrelation gives a diversity gain for SIMO or MISO communications [32], [33]. In the case of MIMO communications, in which the channel model used is characterized by marginal or near-marginal correlation, channel capacity may be greater in the presence of mutual coupling due to the decorrelating effect acting in each array independently [9]. However, this conclusion is limited by the marginal-correlation assumption and is not expected to hold for the more general case of jointly correlated MIMO channels where a joint decorrelation by mutual coupling would be required to improve channel capacity.

The problems posed by coupling-induced correlation and impedance mismatch in antenna arrays suggests that compact MIMO arrays need to be designed with inherently low mutual coupling and good radiation efficiency and that the elements in these arrays should have radiation patterns that interface with all possible multipath angular spectra.

## B. Printed Inverted-F Antennas

A Printed Inverted "F" Antenna (PIFA) has a low profile, good radiation characteristics, and wide bandwidth [34]-[36]. This makes a PIFA an attractive choice for a mobile device antenna in wideband wireless systems. Another requirement of these systems is support for multiband operation. However, standard PIFA designs result in a single band antenna since the antenna's higher order resonant frequencies are not close to its fundamental resonant frequency. In addition, both the radiation performance and the return loss of a standard PIFA antenna may render it inefficient for use at higher resonant frequencies. It has been shown that a PIFA can be made to resonate at a second frequency by using a proper slot on the antenna body or by including a J-shape slot [37]-[39]. A novel tri-band PIFA was designed for use in portable MIMO enabled devices (this design was first used in [19]). The proposed PIFA antenna employs a quarter wavelength slot to provide a second resonant frequency and a J-shape slot to provide a third resonant frequency. The resulting PIFA has resonant frequencies at 2.45 GHz, 5.25 GHz, and 5.8 GHz with supporting bandwidths of 100 MHz, 200 MHz, and 150 MHz respectively. A schematic of a lone PIFA is shown in Fig. 1(a). Fig. 2 shows the measured reflection coefficient of the lone PIFA.

## C. Compact PIFA Arrays

A PIFA is relatively robust to influence from another nearby PIFA because of the radiating element's low profile and proximity to the ground plane. This makes it an ideal candidate for use in a compact array. A 2-element array and a 4-element array were fabricated using tri-band PIFAs. In the remainder of this work, we only present the array characteristics for the 2.45 GHz resonant frequency since the MIMO transceiver testbed into which these arrays were integrated only operates in the 2.4 – 2.5 GHz band.

The fabricated 2-element PIFA array is shown in Fig. 3(b) alongside a laptop PCMCIA wireless card, Fig. 3(d), for size reference. The PIFAs in the 2-element array have a separation of 27 mm ($< \lambda/4$) between their centers. The volume occupied by the array is $0.45\lambda \times 0.12\lambda \times 0.033\lambda$ compared to the PCMCIA card's $0.45\lambda \times 0.25\lambda \times 0.033\lambda$ protruding array housing. Ansoft's *High Frequency Structure Simulator* software package was used to calculate the input impedance as well as the far-field patterns of this array. Fig. 4 shows the computed far-field patterns of a lone PIFA on the ground plane. Fig. 5 shows the far-field patterns of a single PIFA on a ground plane in the presence of a second PIFA located 9 mm away on the ground plane and terminated by a 50 Ω load. The far-field patterns are given for two elevation planes ($\phi_{xz} = 0°$ and $\phi_{yz} = 90°$) and the azimuth plane ($\theta_{xy} = 90°$). A comparison of Fig. 4 and Fig. 5 shows that there is relatively little change in the radiation pattern of a single PIFA due to the presence of a second nearby PIFA.

Fig. 6 presents the measured scattering matrix of the dual element PIFA array as a function of frequency. The results at 2.45 GHz are indicative of an array with excellent return loss and isolation. The difference in return loss between the lone PIFA, Fig 2, and that of the PIFA array indicates how robust this antenna design is to the presence of another nearby antenna. However, it has been shown that the diagonal elements of the scattering matrix, $s_{ii}$, are not a sufficient characterization of radiation efficiency for a multi-port antenna [40]. Firstly, mutual coupling causes some portion of the signal power within each element to be radiated and absorbed by the other elements. This coupling is characterized by the off diagonal elements, $s_{ij}$, of the scattering matrix which are therefore an essential part of a MIMO array characterization. Secondly, the combination of each antenna port's primary reflected signal with the coupled signals can be constructive or destructive depending on the phase of the component signals. Each antenna in a MIMO transceiver is subjected to signals whose harmonic components have some random phase due to transmitter modulation of a random data and phase distortion induced by the propagation environment. MIMO antenna efficiency is therefore a function of how randomly phased signals couple and combine and cannot be determined from the traditional scattering matrix characterization. Instead, an array's Total Active Reflection

Coefficient (TARC) has been defined to give a more meaningful and complete characterization of array efficiency by accounting both for coupling and constructive/destructive signal combining due to random signal phase [40]. TARC is defined as the ratio of the square root of total reflected power divided by the square root of total incident power for a set of random excitation signals. The TARC can be calculated for the $N$-port antenna system with scattering parameter matrix, $\mathbf{S}_P$ as,

$$\Gamma_a^t = \sqrt{\sum_{i=1}^{N}|b_i|^2} \bigg/ \sqrt{\sum_{i=1}^{N}|a_i|^2} \quad , \text{ for } \vec{b} = \mathbf{S}_P \vec{a} \qquad (1)$$

where $\vec{a}$ is the incident signal vector with randomly phased elements and $\vec{b}$ is the reflected signal vector. TARC for an array has the same meaning as the return loss for a single antenna. For example, when 90% of the incident power radiates from the array and 10% is reflected/coupled back to the antenna ports, TARC is equal to $-10\,\text{dB}$. An array's TARC is calculated by applying different combinations of excitation signals to each port. Each of the $N$ excitation signals has unity magnitude but a random initial phase. This random phasing of the excitation signals probes the array in ways that cause anything from constructive to destructive combining of coupled signals with the primary reflected signal. This excitation method more accurately simulates MIMO signaling than do the signals traditionally used in scattering matrix characterization. Note that signals combine in the linear scale (voltages) whereas the resulting power signal, used in computing TARC, is the squared magnitude of the voltage signals. Therefore, TARC can increase rapidly as the number of array elements increases. The TARC of the 2-element PIFA array was calculated for a set of twenty excitation vectors and the results are presented in Fig. 7. The worst case calculated TARC is lower than $-9$ dB for different excitation vector combinations. This result is indicative of an array with good efficiency across a range of possible MIMO signals.

The design for the 2-element PIFA array was adapted to construct a 4-element array by using both sides of the ground plane. A schematic of the four element PIFA array is shown in Fig. 1(b). Fig. 3(c) shows the top and bottom views of the manufactured 4-element PIFA array. Two PIFAs are mounted above the ground plane and two are mounted directly below so that those above are mirrored across the ground plane by those below. PIFAs on the same side of the ground plane mirror each other as in the 2-element PIFA array. Two RT-Duroid substrates are attached back-to-back to realize four microstrip lines to feed the array's elements. The PIFAs in the 4-element array have a maximum separation of 28 mm ($< \lambda/4$) between their centers and a minimum separation of 8 mm ($< \lambda/16$). The volume occupied by the 4-element PIFA array measures $0.45\lambda \times 0.12\lambda \times 0.066\lambda$ and is believed to be the smallest yet reported for a 4-element array in its class. Due to the small distance between top and bottom elements, a higher level of mutual coupling can be expected between element pair $(1,3)$ and $(2,4)$. Fig. 8 shows one row of the scattering matrix of the 4-element PIFA array. A marked but acceptable degradation is seen when comparing with the 2-element array's return loss and isolation. However, as with the 2-element array, the scattering matrix characterization is not a sufficient indicator of MIMO performance. The TARC of the 4-element PIFA array was calculated and the result is presented in Fig. 9. We see in comparing the TARC in Fig. 7 and Fig. 9 that TARC increases rapidly with increasing number of elements. This resulting loss of efficiency is an expected consequence of the increasing mutual coupling as antenna density increases. This also demonstrates how much more difficult it is to realize a compact array with four elements than it is to realize one with just two elements.

*D. Reference Dipole Arrays*

A ULA of dipole antennas was used as a baseline array for providing a performance reference. These dipoles, shown in Fig. 3(a), have resonant frequencies, polarization and peak gain that closely match those of the PIFA. Each dipole measures $133.5\,\text{mm} \times 17.6\,\text{mm} \times 6\,\text{mm}$ (or $1.1\lambda \times 0.15\lambda \times 0.05\lambda$). Dipole arrays with antenna spacings of $\Delta_d \in \{\lambda, \lambda/2, \lambda/4, \lambda/8\}$ are used in this study. Two of these arrays are shown in Fig. 3(e)(f). The volume of a $\Delta_d = \lambda/2$ dipole ULA, $V_d = (17.6\,\text{mm} \times 3(60.2\,\text{mm}) \times 133.5\,\text{mm})$, is 75 times greater than that of the 4-element PIFA array's volume $V_{\text{4-PIFA}} = (15\,\text{mm} \times (2(9\,\text{mm}) + 27\,\text{mm}) \times 2(4\,\text{mm}))$. The scattering parameters for the dipole arrays were measured for each antenna spacing. A subset of the s-parameter matrix for the dipole arrays is presented in Table 1. The dipole array scattering parameter are seen to generally degrade as $\Delta_d$ decreases. This indicates the loss of radiation efficiency as the array aperture shrinks.

## III. MIMO Performance Characterization

The second general contribution of this study is the characterization of compact MIMO array performance in field trials. This section addresses the theoretical tools and experimental methodology that allow a fair and meaningful characterization of MIMO arrays. Channel capacity is the quintessential measure of MIMO system performance. The necessary theoretical framework for computing ergodic MIMO capacity is presented first. The computation of ensemble average MIMO capacity from field measurements is considered next. Normalization of measured capacity is a critical issue when comparing measurements made with different arrays over a variety of propagation scenarios. This issue is addressed with particular care. An appropriate method for characterizing the MIMO channel's measured eigenvalue distributions is then presented. This characterization gives an indication of how well an antenna array's radiation pattern interfaces with multipath propagation. Finally, a description is given of an extensive MIMO field measurement campaign using a true-MIMO testbed designed to characterize MIMO array performance in a range of typical indoor communications scenarios.

## A. MIMO System Model

A narrowband MIMO communications system is modeled as,

$$\vec{Y} = \mathbf{H}\vec{X} + \vec{\mathbf{N}}, \quad (2)$$

where $\vec{Y} \in \mathbb{C}^{L_R \times 1}$ is the received signal vector, $\mathbf{H} \in \mathbb{C}^{L_R \times L_T}$ is the qausi-static MIMO channel distortion matrix, $\vec{X} \in \mathbb{C}^{L_T \times 1}$ is the transmitted signal vector, and $\vec{\mathbf{N}} \in \mathbb{C}^{L_R \times 1}$ is the additive noise vector introduced at the receiver. The following assumptions are made in order to compute channel capacity for this model:

i) $\mathbf{H}$ is a matrix of i.i.d. circularly symmetric zero mean Gaussian random variables, $\mathbf{h}_{ij}$, with variance $\sigma_\mathbf{h}^2$.

ii) $\vec{\mathbf{N}}$ is a vector of i.i.d. zero mean Gaussian random variables, $\mathbf{n}$, with variance $\sigma_\mathbf{n}^2$.

iii) The transmitter has *a priori* knowledge of channel statistics and the receiver knows $\mathbf{H}$.

iv) The total transmit power, $P_\Sigma$, is allocated uniformly to each transmit antenna as $P = P_\Sigma/L_T$.

Given this model and these assumptions, the average ergodic MIMO channel capacity is computed as [1],

$$\mathbf{C}(\sigma_\mathbf{h}^2, \sigma_\mathbf{n}^2) = \mathcal{E}_\mathbf{H}\left\{\log_2 \det\left(\mathrm{I}_{L_R} + \frac{P}{\sigma_\mathbf{n}^2}\mathbf{H}\mathbf{H}^\mathrm{H}\right)\right\}, \quad \text{(b/s/Hz)} \quad (3)$$

where $\mathcal{E}_\mathbf{H}\{\,\cdot\,\}$ is the statistical expectation taken over $\mathbf{H}$. The average Signal to Noise Ratio (SNR) per receiver is,

$$\rho = P_\Sigma \sigma_\mathbf{h}^2 / \sigma_\mathbf{n}^2. \quad (4)$$

## B. Computing MIMO Capacity

A complementary framework to the system model above has to be defined in order to compute MIMO capacity from wideband field measurements. Let the channel at frequency, $f_l$, and position, $s_m$, be $\mathcal{H}(f_l, s_m) \in \mathbb{C}^{L_R \times L_T}$ and time-invariant during the period of observation. We occasionally use the abbreviated notation, $\mathcal{H}$, to represent $\mathcal{H}(f_l, s_m)$ and note that $\mathcal{H}$ is a realization of the random matrix $\mathbf{H}$. For fixed transmit power and fixed receiver noise variance, the instantaneous channel capacity, $C(\mathcal{H}, \sigma_\mathbf{n}^2)$, is computed as,

$$C(\mathcal{H}, \sigma_\mathbf{n}^2) = \log_2 \det\left(\mathrm{I}_{L_R} + \frac{P}{\sigma_\mathbf{n}^2}\mathcal{H}\mathcal{H}^\mathrm{H}\right), \quad (5)$$

Since we are interested in ergodic capacity, $\mathbf{C}(\sigma_\mathbf{h}^2, \sigma_\mathbf{n}^2)$, we can estimate it by computing the ensemble average channel capacity, $\overline{C(\mathcal{H}, \sigma_\mathbf{n}^2)}$, of the measured channels as follows. From a measurement at position $s_m$ and frequency $f_l$, we compute an estimate of the channel, $\widehat{\mathcal{H}}(f_l, s_m)$. This is repeated over $K$ uniformly spaced frequencies in the measurement band and over a uniform grid of $M$ spatial position-pairs of the transmitter and receiver arrays. An estimate of the receiver noise variance, $\widehat{\sigma}_{\mathbf{n}}^2$, is computed at the same time as the channel estimates. An appropriate method for estimating the channel coefficients and noise is given in [41]. The ensemble average channel capacity is finally computed as,

$$\overline{C(\widehat{\mathcal{H}},\widehat{\sigma}_{\mathbf{n}}^2)} = \frac{1}{KM} \sum_{m=1}^{M} \sum_{k=1}^{K} \log_2 \det\left(\mathbf{I}_{L_R} + \frac{P}{\widehat{\sigma}_{\mathbf{n}}^2} \widehat{\mathcal{H}}(f_k,s_m)\widehat{\mathcal{H}}^{\mathrm{H}}(f_k,s_m)\right), \qquad (6)$$

The capacity computed in (6) is sensitive to channel correlation and array radiation efficiency by way of their effect on $\widehat{\mathcal{H}}(f_k,s_m)$. It is critical that $\{\widehat{\mathcal{H}}(f_k,s_m)\}$ not be normalized, as is typically done in MIMO channel sounding studies, since the effect that the arrays' antenna interactions would then be hidden by the normalization. An appropriate normalization method is discussed in Section III-C.

Capacity of MIMO systems with dimensions less than $L_T \times L_R$ is computed from $L_T \times L_R$ measurements by choosing a subset of the elements of $\widehat{\mathcal{H}}$ and reallocating power in the transmit antenna to keep the total transmit power constant. For example, a $4 \times 4$ measurement of $\widehat{\mathcal{H}}$ is used to compute the channel capacity for the $3 \times 3$ case by choosing the submatrix of $\widehat{\mathcal{H}}$ comprised of the elements, $\widehat{h}_{(i,j)}$, where $i,j \leq 3$. The transmit power allocation is then recomputed, according to $P = P_\Sigma/3$, so that a fair comparison of channel capacity can be made between the $4 \times 4$ and $3 \times 3$ cases.

## C. Normalizing MIMO Capacity

A comparison of the MIMO capacity achieved by each type of array must be made using measurements from a variety of sites. To make a fair comparison, measured ensemble average channel capacity will be normalized as follows. A reference SNR measurement is made at each site using dipole ULAs with $\Delta_d = \lambda$ antenna spacings. Mutual coupling is assumed to have negligible effect on receiver SNR for such large spacings. The reference measurement is made using the same grid positions, $\{s_m\}$, and frequencies, $\{f_l\}$, used for all site measurements. A reference noise variance, $\sigma_{\mathbf{n},R}^2$, is computed as before while the reference channel variance is computed as,

$$\sigma_{\mathbf{h},R}^2 = \frac{1}{KML_TL_R} \sum_{m=1}^{M} \sum_{k=1}^{K} \mathrm{vec}(\widehat{\mathcal{H}}(f_k,s_m))^H \mathrm{vec}(\widehat{\mathcal{H}}(f_k,s_m)). \qquad (7)$$

The normalized ensemble average channel capacity is then computed as,

$$\overline{C_N} = \frac{\overline{C(\widehat{\mathcal{H}},\widehat{\sigma}_{\mathbf{n}}^2)}}{\mathbf{C}(\sigma_{\mathbf{h},R}^2,\sigma_{\mathbf{n},R}^2)} \qquad (8)$$

where $\mathbf{C}(\sigma_{\mathbf{h},R}^2,\sigma_{\mathbf{n},R}^2)$ is the i.i.d. Gaussian channel capacity corresponding to the reference SNR measurement and is computed by Monte Carlo simulation of the system defined in Section III-A. The computed reference channel capacity is assumed to be what could be achieved in the best case scenario of no mutual coupling and no channel correlation. This normalization is appropriate in the high SNR regime where MIMO capacity is dominated by the $(P/\sigma_\mathbf{n}^2)\mathbf{HH}^\mathrm{H}$ term in (3). The normalization removes site-specific SNR bias caused by large scale fading but does not hide the effect that mutual coupling has on channel capacity. This capacity normalization is a departure from the traditional method in which the measured $\{\widehat{\mathcal{H}}(f_k, s_m)\}$ are normalized and SNR is set to a constant before computing capacity. Normalizing the channel before computing capacity masks the effect that mutual coupling has on receiver SNR and MIMO channel capacity [11].

## D. Eigen-Channel Characterization

The eigen decomposition of the Hermitian symmetric matrix, $\mathcal{H}\mathcal{H}^\mathrm{H}$, yields,

$$\mathcal{H}\mathcal{H}^\mathrm{H} = U\Lambda U^\mathrm{H}, \tag{9}$$

where $U \in \mathbb{C}^{L_R \times 1}$ is a unitary matrix whose columns are the eigenvectors of $\mathcal{H}\mathcal{H}^\mathrm{H}$ and $\Lambda$ is a diagonal matrix of the eigenvalues, $\{\lambda_1, \lambda_2, ..., \lambda_{L_\lambda}\}$ of $\mathcal{H}\mathcal{H}^\mathrm{H}$. Each eigenvector/eigenvalue pair describes a MIMO eigen-channel. If both the transmitter and receiver know $\mathcal{H}$, then the instantaneous MIMO capacity can be computed in terms of the eigenvalues of $\mathcal{H}\mathcal{H}^\mathrm{H}$ as,

$$C(\mathcal{H}, \sigma_\mathrm{n}^2) = \sum_{i=1}^{L_\lambda} \log_2\left(1 + \frac{P}{\sigma_\mathrm{n}^2}\lambda_i(f_k, s_m)\right), \tag{10}$$

were $L_\lambda = \min(L_T, L_R)$ denotes the number of non-zero eigenvalues of $\mathcal{H}\mathcal{H}^\mathrm{H}$. The relationship between the eigenvalues and channel capacity is most clearly seen in (10). Each eigenvalue is a measure of the strength of the associated eigen-channel. The orthoganality of the eigen decomposition is what enables each of these eigen-channels to support independent information streams.

A comparison of the eigenvalue probability distributions measured with different arrays gives an indication of how well each array preserves independence in the MIMO subchannels. Such comparisons can be made between different arrays for the same measurement site in a way that reflects how well each array's radiation pattern interfaces with the propagation mutlipath while discounting differences between array efficiency (which only affects capacity via SNR). How this is done is described as follows. Eigen decompositions are computed for a channel measurement set, $\{\widehat{\mathcal{H}}(f_k, s_m)\}$, associated with a particular type of array. The eigenvalues for each $\widehat{\mathcal{H}}(f_k, s_m)$ are ordered such that $\widehat{\lambda}_1 \geq \cdots \geq \widehat{\lambda}_{L_\lambda} > 0$. The average channel gain is computed for the measurement set according to,

$$G = \frac{1}{MKL_\lambda} \sum_{m=1}^{M} \sum_{l=1}^{K} \sum_{i=1}^{L_\lambda} \widehat{\lambda}_i(f_k, s_m), \qquad (11)$$

This is repeated for measurement sets associated with each type of array. The eigenvalue distributions for an i.i.d. Gaussian MIMO channel are computed by Monte Carlo simulation of the system described in Section III-A with $\sigma_{\mathbf{h}}^2 = 1$. The resulting average channel gain of the i.i.d. Gaussian MIMO channel is $G_{\text{i.i.d.}} = L_T L_R$. The eigenvalue measurement set, $\{\widehat{\lambda}_i(f_k, s_m)\}$, associated with each array is then scaled so that $G = G_{\text{i.i.d.}}$. The measured distributions of the ordered eigenvalues, $\{P(\widehat{\lambda}_1), ..., P(\widehat{\lambda}_{L_\lambda})\}$ are then estimated by histogram. This normalization removes the effect that large scale fading and impedance mismatch have on SNR but does not hide the effect that propagation-induced correlation or coupling-induced correlation have on the channel's eigenvalue distributions.

## E. Field Measurement Procedure

Over-the-air measurements were made with a true-MIMO radio testbed that broadcasts a 16 MHz wideband excitation signal from all transmit antenna and simultaneously captures the received signal on all receiver antenna [42], [43]. The channel's coherence bandwidth at this frequency is less than 16 MHz for typical indoor environments. A Space-Frequency Orthogonal Multitone (SFOM) channel sounding scheme was used to measure $\widehat{\mathcal{H}}(f_l, s_m)$ and $\widehat{\sigma}_N^2$ [41]. A complete calibration of the each radio's gain, phase noise, complex spectrum and frequency offset was performed prior to the campaign. Benchmark tests on an Elektrobit C8 MIMO channel emulator were performed to verify the testbed's calibration. These benchmarks included channel capacity measurements for a $2 \times 2$ i.i.d. Gaussian channel and a $2 \times 2$ perfectly correlated Gaussian channel and produced results consistent with the ergodic channel capacity expected for i.i.d. Gaussian MIMO and Gaussian SISO channels respectively.

The measurement sites were located in a multi-story building whose floorplan included laboratories, classrooms, offices and corridors. Two measurement sites were chosen: (i) a laboratory with electronic equipment, cubicle partitions and office furniture; (ii) a very large empty hall with a few support columns and no furnishings. At each site, a Line-of-Sight (LOS) and a Non-Line-of-Sight (NLOS) visibility scenario was chosen for placing the transmitter (TX) and receiver (RX) arrays.

LEGO positioners were used to move the transmitter and receiver antenna arrays simultaneously and parallel to the plane of the floor in a pattern defined by a $50\,\text{cm} \times 60\,\text{cm}$ grid of $11 \times 9$ points. Antenna arrays were mounted on the positioners at a height of 1.3m above the floor. The relative angle between the TX and RX array endfires ranged between 20° and 70° at each site. PIFA arrays were mounted with the ground plane parallel to the plane of the floor. Dipole arrays were mounted with the dipoles perpendicular to the plane of the floor. The TX and RX radios were outfitted with a particular TX-RX pair of arrays, for example array $A$ and array $B$. Measurements were taken first with the (Node 1, Node 2) array-pair being $(A, B)$ and then repeated with the array-pair being $(B, A)$. In this way, a total of $K \times M \times 2 = 9504$ narrowband channel matrices were recorded for each array-pair at each test site. Other measurement parameters are presented in Table 2.

## IV. Field Trial Results

This section presents field trial results from an extensive measurement campaign that was designed to characterize the performance of compact MIMO arrays in typical indoor environments. Measured channel capacity results are first presented from one representative measurement site to show the effects that impedance mismatch and channel correlation have on SNR and channel capacity. Next, measured eigenvalue distributions from the same representative scenario are presented to show the effect that a compact array's radiation pattern has on MIMO eigen-channel structure. Finally, a comprehensive set of results are presented from all measurement locations and a fair comparison of various compact MIMO arrays' performances is made.

### A. Channel Capacity vs. Array Aperture

Measurements involving dipole ULAs showed how strongly SNR and channel capacity vary as a function antenna separation. As an example, Fig. 10 presents the measured MIMO channel capacity for three different combinations of ULA antenna separation in the Empty Hall NLOS scenario. Each datum point marks the average narrowband channel capacity computed according to the inner sum in (6) at one grid position in the $11 \times 9$ grid of spatial samples. The ergodic MIMO capacities of i.i.d. Gaussian channels, given by (3), are plotted for reference. For the $4 \times 4$ case, ensemble average capacity drops from $33.5\,\text{b/s/Hz}$ to $27.0\,\text{b/s/Hz}$ to $22.3\,\text{b/s/Hz}$ as first one array and then the other collapses from $\Delta_d = \lambda$ to $\Delta_d = \lambda/8$. Average SNR drops simultaneously from $29.9\,\text{dB}$ to $28.6\,\text{dB}$ to $27.1\,\text{dB}$. The slight decrease in SNR is a symptom of increasing impedance mismatch between the arrays and the $50\Omega$ radios. SNR degradation accounts for only $4\,\text{b/s/Hz}$ of the total $11.2\,\text{b/s/Hz}$ capacity lost while the remainder is attributed to increasing channel correlation. These trends were observed without exception for all measurement sites and demonstrates the dominating effect that correlation has on MIMO capacity. While this observation is not unexpected, these are the first true-MIMO field results to show the extent of the effect that correlation and SNR have on channel capacity for very small antenna spacings. Interestingly, the $2 \times 2$ case proves to be far more robust to shrinking array apertures. As mentioned in the discussion of TARC in Section II-C, this field result is confirmation that mutual coupling becomes an increasingly important problem as the number of antenna increase. A consequence of this result is that MIMO capacity does not scale linearly as $\mathcal{O}(\min(L_T, L_R))$ in the presence of strong mutual coupling. It has previously been shown that MIMO channel capacity grows linearly in the number of antennas, even in the presence of propagation-induced correlation [44]. It is therefore necessary to consider both coupling-induced correlation and propagation-induced correlation when considering MIMO channel capacity in real world transceiver implementations.

## B. Channel Eigenvalue Distributions

Degradation in channel capacity due to shrinking array aperture shows the gross effects of array efficiency and channel correlation. The more subtle relationship between channel capacity and channel correlation can be observed from the effect that a shrinking array aperture has on the MIMO channel's eigenvalue distribution. As an example, Fig. 11 presents the measured ordered eigenvalue distributions for two different ULA apertures and for the 4-element PIFA array in the Empty Hall NLOS scenario. These distributions are normalized (a left shifting of the measured distributions on the figure's dB scale) as discussed in Section III-D so that the measured channel gain matches that of a reference i.i.d Gaussian channel. Each measured distributions are generated from a histogram of 4752 channel matrices over a 60 dB bin domain with a 0.5 dB bin spacing. The i.i.d. Gaussian MIMO channel's ordered eigenvalue distributions are shown for reference.

A comparison of Fig. 11(a) and 11(b) reveals that the strongest eigenvalue becomes stronger and the remaining eigenvalues become weaker as the dipole array collapses from $\Delta_d = \lambda$ to $\Delta_d = \lambda/8$. The channel is seen to collapse from an excellent MIMO channel to something resembling more of a SISO or beamformed channel. It is clear from (10) that the logarithmic increase in capacity due to increasing one eigen-channel does not compensate for the decrease in capacity due to a balanced weakening of the other eigen-channels. In comparison, Fig. 11(c) shows results from the same measurement but where a 4-element PIFA array has replaced the collapsed dipole ULA. Unlike the collapsed ULA element's more directional radiation patterns [33], the PIFA array's elements have relatively non-directional radiation patterns (see Fig. 5). The PIFA array's resulting eigenvalue distributions are nearly as close to the ideal i.i.d. Gaussian channel's distributions as those of the $1\lambda$ dipole ULA, despite the fact that the PIFA array is subject to the same or worse propagation-induced correlation. This is a critically important measurement result because it shows that MIMO arrays with directional radiation patterns will induce correlation in a channel that is otherwise i.i.d. Gaussian. The consequence of this forced directionality in the channel prevents full MIMO capacity from being achieved because of how power is redistributed to the channel's strongest eigenvalue (though this capacity be still greater than that possible for the same SNR using single antennas). The fact that the eigen-channel structure perceived using the PIFA array is nearly the same as that of an i.i.d. Gaussian channel demonstrates how well the PIFA array element's radiation patterns interface with the available multipath propagation and is one reason that the PIFA array is better at preserving the inherent MIMO channel capacity as will be seen in the next section. Again, these trends were observed without exception for all measurement sites.

## C. Performance Comparison of Arrays

Comprehensive results of MIMO array performance are presented in Fig. 12. For a particular measurement site, the ensemble average capacity associated with each type of array-pair (see Section III-B) was normalized according to the procedure described in Section III-C. These normalized average capacities were then averaged over all sites so that each datum represents the normalized ensemble average capacity of approximately $38,000$ narrowband MIMO channels. These data are therefore assumed to give a statistically significant characterization of each array's performance in a typical indoor environment. Note that the operating SNR during all measurements was sufficiently high to be considered in the linear region of the i.i.d. Gaussian capacity curves (e.g.. see Fig. 10). The high SNR not only allows good channel estimates to be computed but also the near linear relationship between SNR and capacity makes the capacity normalization assumptions in Section III-C appropriate.

The combined plots from Fig. 12(a) and Fig. 12(b) are a continuum of shrinking first one node's array aperture and then shrinking the other node's array aperture Indeed, the right-most data in the Fig. 12(a) are the same as the left-most data in Fig. 12(b). Note also that the data at $\Delta_d^{(N1)}/\lambda = 1$ are the reference measurements described in Section III-B whose SNRs are used to compute the i.i.d. Gaussian capacities for normalizing the data in Fig. 12. Fig. 12(b) is discussed first because of the natural left-to-right flow of the shrinking array apertures suggested by the side-by-side plots.

Fig. 12(b) shows how MIMO channel capacity is affected by a link where one node has a shrinking array aperture and the other node has a large fixed array aperture This scenario is relevant to indoor MIMO system deployments where a base station can support large arrays $(\Delta_d \geq \lambda/2)$ and the mobile devices must use compact arrays $(\Delta_d < \lambda/2)$. The capacity achieved by replacing the Node 1's collapsed dipole ULA with 2-element and 4-element PIFA arrays are shown at $\Delta_d^{(N1)}/\lambda = 1/8$. The 2-element PIFA array in a $2 \times 2$ MIMO configuration outperforms the smallest dipole array by $5.4\%$. The 4-element PIFA array in $2 \times 2$, $3 \times 3$ and $4 \times 4$ MIMO configurations outperforms the smallest dipole array by $3.4\%$, $7.3\%$ and $7.9\%$ respectively. The PIFA arrays are seen to have an equivalent dipole ULA spacing in the $\lambda/4$ to $\lambda/2$ range. However, it would seem from a comparison of the 2-element PIFA array's scattering parameters with those of the dipole ULAs in, both given Table 1, that a dipole array with $\Delta_d = \lambda/4$ would outperform the PIFA array. This is a clear example of why a scattering matrix measurement does not sufficiently characterize an array's performance under MIMO signaling conditions.

Fig. 12(a) shows how MIMO channel capacity is affected by a link where one node has a shrinking array aperture and the other node has a compact array aperture This scenario examines how robust various compact mobile arrays are to an array at the other node with decreasing performance. The 2-element PIFA is clearly the most robust and shows only a $4.4\%$ degradation compared to $\approx 10\%$ degradation for the 4-PIFA array and dipole ULAs. The 2-element PIFA array's robustness is consistent with results presented in [19]. As discussed in Sections II-C and IV-B, the 4-element PIFA suffers more from mutual coupling than the 2-element PIFA. Even so, the 4-element PIFA is seen to give consistently more robust performance than the larger dipole ULA.

## V. Conclusion

This work has presented a study of compact antenna arrays for MIMO radio communications with two general contributions: (i) the realization of compact arrays that are designed to preserve MIMO channel capacity without the need for matching networks and (ii) the characterization of compact array performance by field measurements using MIMO radios.

A novel triband PIFA was designed to have a low profile, good radiation characteristics, and wide bandwidth. This PIFA design proved to be highly suited to compact MIMO arrays because of its robustness to the influence of another nearby PIFA element. Two highly compact antenna arrays were designed and fabricated for use in MIMO enabled mobile devices. Instead of only performing a scattering matrix characterization of the compact arrays, a Total Active Reflection Coefficient characterization was used because it's ability to measure the effect of mutual coupling on array efficiency under MIMO signaling conditions. The 2-element PIFA array demonstrated excellent isolation and efficiency while the 4-element PIFA array demonstrated good performance while having the smallest volume yet reported for a 4-element array in its class.

An experimental methodology was formulated to allow a fair and meaningful characterization of MIMO arrays by field trial. This methodology addressed the issue of capacity normalization in a way that did not mask the effects of mutual coupling but still allowed a fair comparison of antenna array performance from measurements taken in different propagation scenarios. A statistical method was also presented for quantifying how well an antenna array's radiation pattern interfaces with multipath propagation by characterizing the MIMO channel's underlying eigenstructure.

This experimental methodology was implemented in a measurement campaign in which compact PIFA arrays and dipole arrays of various size were evaluated for their ability to preserve MIMO channel capacity. Measurements were made using a true-MIMO transceiver testbed that captured the effect of mutual coupling in the reported capacity measurements. Measurement results showed the extent to which impedance mismatch and channel correlation affect channel capacity as the array aperture shrinks. It was observed that MIMO capacity does not scale linearly with the number of antenna elements in the presence of strong mutual coupling. This result is distinct from the linear capacity scaling that is possible in the presence of propagation-induced correlation. A study of the measured eigenvalue distributions showed that, unless the compact array's elements had a nearly omnidirectional pattern, the array induced correlation in a channel that was otherwise i.i.d. Gaussian. Forced directionality in the perceived channel prevented full MIMO capacity from being achieved because of how the channel's eigenvalues became redistributed to have a stronger SISO characteristic. Finally, a comprehensive comparison of the channel capacity achieved over all measurement sites showed that the compact PIFA arrays performed as well as much larger dipole arrays with antenna spacings in $\lambda/4$ to $\lambda/2$ range. The 2-element PIFA array was observed to have an exceptional ability to preserve MIMO channel capacity, even when a very poorly performing array was used at the other end of a $2 \times 2$ MIMO link. This result was reinforced by TARC calculations which demonstrated that mutual coupling becomes an increasingly important problem as the number of antennas increase, even when the increase is from two antennas to four. The compact array's good performance was achieved without the use of matching networks. The difficulty involved in realizing wideband and multiband matching networks as compact circuits makes the reliability of compact arrays with inherently acceptable mutual coupling and radiation efficiency, as demonstrated in this work, a very attractive alternative.

ACKNOWLEDGMENT

The Authors would like to thank Michael Parker at the UCLA Department of Computer Science for developing software for the positioning robots.REFERENCES

[1] I. E. Telatar, "Capacity of Multi-Antenna Gaussian Channels," *AT&T Bell Lab. Tech.* Memo, June 1995.

[2] G. Foschini, and M. Gans, "On limits of wireless communications in a fading environment when using multiple antennas," *Wireless Personal Communications*, vol. 6, pp. 311- 335, 1998.

[3] D. S. Shiu, G. J. Foschini, M. J. Gans, J. M. Kahn, "Fading correlation and its effect on the capacity of multielement antenna systems," *IEEE Trans. Commun.*, vol. 48, pp. 502-513, 2000.

[4] D. Chizhik, J. Ling, P. Wolniansky, R. Valenzuela, N. Costa, K. Huber, "Multiple-input-multiple-output measurements and modeling in Manhattan," *IEEE Journal on Selected Areas in Communications*, vol. 21 , pp. 321 - 331, 2003 .

[5] C. A. Balanis, *Antenna Theory: Analysis and Design*, Wiley, 1997.

[6] W. Lee, "Effects on Correlation Between Two Mobile Radio Base-Station Antennas," *IEEE Transactions on Communications*, vol. 21, pp. 1214 - 1224, 1973.

[7] I. Gupta and A. Ksienski, "Effect of mutual coupling on the performance of adaptive arrays," *IEEE Transactions on Antennas and Propagation*, vol. 31, no.5, pp. 785- 791, Sep 1983.

[8] T. Svantesson and A. Ranheim, "Mutual coupling effects on the capacity of multielement antenna systems," *in Proc. IEEE Int. Conf. Acoustics, Speach, and Signal Processing*, vol. 4, pp. 2485–2488, May, 2001.

[9] R. Janaswamy, "Effect of element mutual coupling on the capacity of fixed length linear arrays," *IEEE Antennas Wireless Propagat. Lett.*, vol. 1, no. 1, pp. 157–160, 2002.

[10] J. W. Wallace and M. A. Jensen, "Mutual coupling in MIMO wireless systems: a rigorous network theory analysis," *IEEE Transactions on Wireless Communications*, vol. 3, pp. 1317 - 1325, 2004.

[11] C. Waldschmidt, S. Schulteis, W. Wiesbeck, "Complete RF system model for analysis of compact MIMO arrays," *IEEE Transactions on Vehicular Technology*, vol. 53, pp. 579 - 586, 2004.

[12] W. C. Y. Lee, "Mutual Coupling Effect on Maximum-Ratio Diversity Combiners and Application to Mobile Radio," *IEEE Transactions on Communications*, vol. 18, pp. 779 - 791, 1970.

[13] J. W. Wallace and M. A. Jensen, "Termination-dependent diversity performance of coupled antennas: Network theory analysis," *IEEE Trans. Antennas Propagat.*, vol. 52, pp. 98–105, Jan. 2004.

[14] A. Grau, J. Romeu, F. De Flaviis, "On the Diversity Gain Using a Butler Matrix in Fading MIMO Environments," *IEEE International Conference on Wireless Communications and Applied Computational Electromagnetics*, pp. 478 - 481, April 2005.

[15] M. A. Jensen and Y. Rahmat-Samii, "FDTD analysis of PIFA diversity antennas on a hand-held transceiver unit," in *IEEE Antennas Propagat. Symp. Dig.*, pp. 814-817, June 1993.

[16] J. S. Colburn, Y. Rahmat-Samii, M. A. Jensen, G. J. Pottie, "Evaluation of personal communications dual-antenna handset diversity performance," *IEEE Transactions on Vehicular Technology*, vol. 47, no. 3, pp. 737 - 746, Aug. 1998.

[17] M. Karaboikis, C. Soras, G. Tsachtsiris, V. Makios, "Compact dual-printed inverted-F antenna diversity systems for portable wireless devices." *IEEE Antennas and Wireless Propagation Letters*, vol. 3, pp. 9 - 14, 2004.

[18] Y. Gao, C. C. Chiau, X. Chen, C. Parini, "Modified PIFA and its array for MIMO terminals," *IEE Proceedings on Microwaves, Antennas and Propagation*, vol. 152, pp. 255- 259, August 2005.

[19] D. W. Browne, M. Manteghi, M. P. Fitz, Y. Rahmat-Samii, "Antenna Topology Impacts on Measured MIMO Capacity," *IEEE Vehicular Technology Conference*, September 2005.

[20] R. R. Ramirez, F. De Flaviis, "A mutual coupling study of linear and circular polarized microstrip antennas for diversity wireless systems," *IEEE Transactions on Antennas and Propagation*, vol. 51, no. 2, pp. 238 - 248, Feb. 2003

[21] B. A. Cetiner, L. Jofre, J. Y. Qian, F. De Flaviis, "Small-size broadband multi-element antenna for RF/wireless systems," *Antennas and Wireless Propagation Letters* , vol.2, pp. 326 - 329, 2003.18

# TABLES AND FIGURES

**TABLE 1**
**Measured Scattering Parameters of Dual Element Arrays at 2.49 GHz**

|  | $s_{11}$ (dB) | $s_{12}$ (dB) |
|---|---|---|
| Dipole Array |  |  |
| $\Delta_d = \lambda$ | $-25.6$ | $-22.3$ |
| $\Delta_d = \lambda/2$ | $-24.1$ | $-20.1$ |
| $\Delta_d = \lambda/4$ | $-26.2$ | $-15.8$ |
| $\Delta_d = \lambda/8$ | $-17.8$ | $-11.0$ |
| 2-PIFA Array ($\Delta_d < \lambda/4$) | $-20.5$ | $-14.3$ |

**TABLE 2**
**Measurement Campaign Parameters**

| Site | Visibility | TX-RX Separation |
|---|---|---|
| lab | LoS | 10 m |
| lab | NLoS | 10 m |
| empty hall | LoS | 13 m |
| empty hall | NLoS | 28 m |

| | |
|---|---|
| Center Frequency | 2.49 GHz |
| Signal Bandwidth | 16 MHz |
| Signal Power | 15 dBm per TX |
| Signaling Scheme | SFOM true-MIMO |
| # Tones per TX | $K = 48$ |
| Positioner Grid | $M = 11 \times 9$ |
| Gridpoint Spacing | $\lambda/2$ |

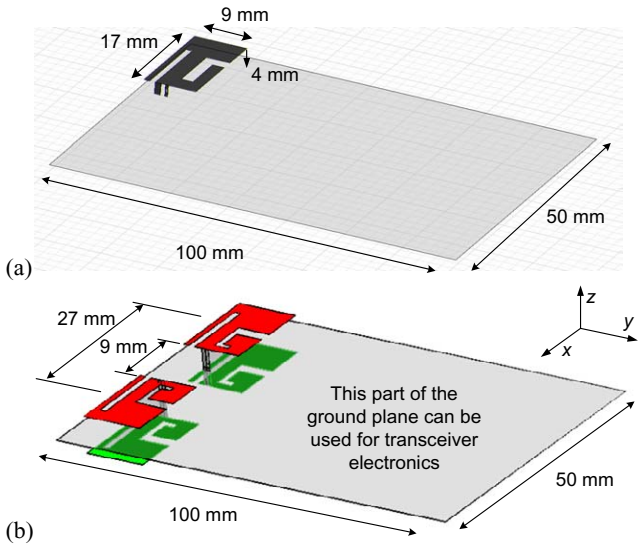

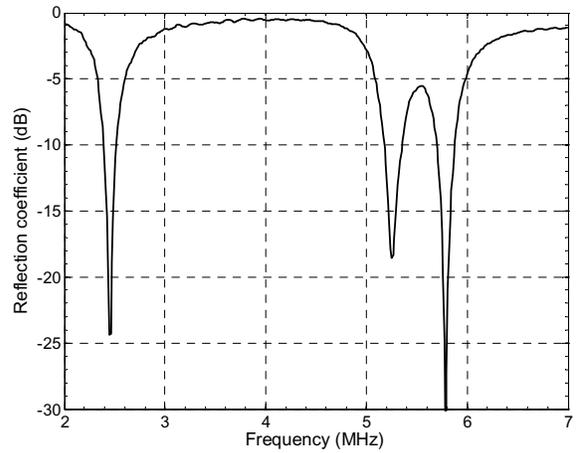

Fig. 1: Antenna schematics of (a) a lone tri-band PIFA element on a ground plane, (b) a 4-element PIFA array with two PIFA above the ground plane and two PIFA below it.

Fig. 2: Measured reflection coefficient of a lone PIFA on the groundplane.

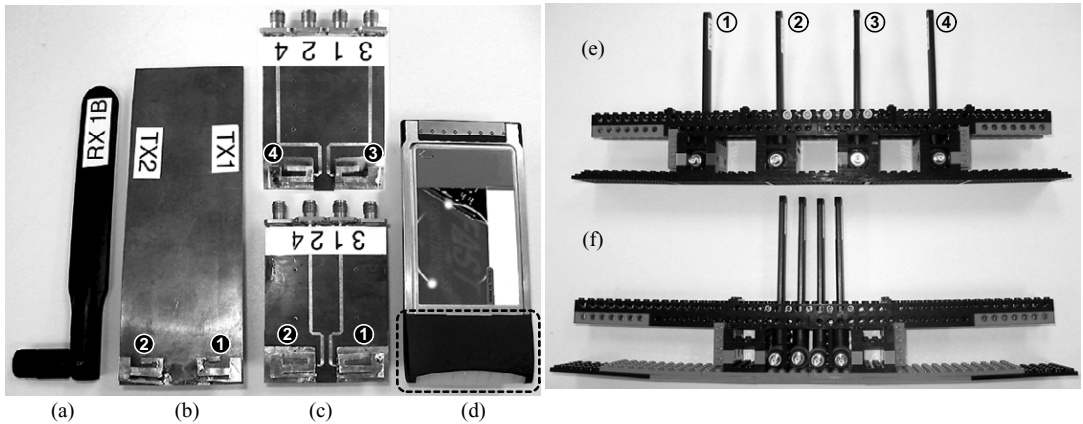

Fig. 3: Antenna used in the measurement campaign: (a) dipole, (b) 2-element PIFA array, (c) 4-element PIFA array from below and above, (d) commercial PCMCIA wireless card with diversity array housing circled for size reference, (e) dipole ULA of the type shown in (a) with $\Delta_d = \lambda/2$, (f) dipole ULA of the type shown in (a) with $\Delta_d = \lambda/8$.

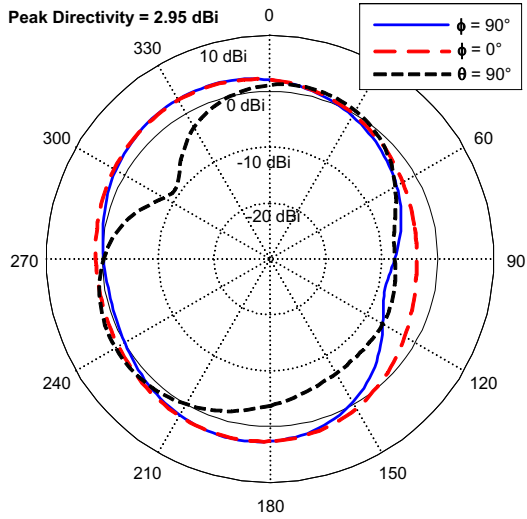

Fig. 4: Calculated far field azimuth and elevation patterns at 2.45 GHz for a lone PIFA on the groundplane.

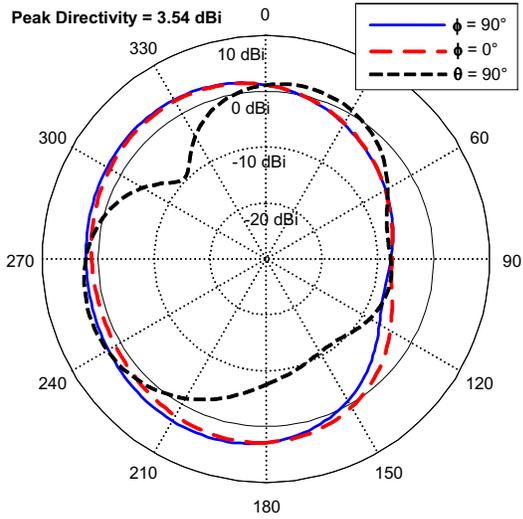

Fig. 5: Calculated far field azimuth and elevation patterns at 2.45 GHz for a single PIFA in the dual PIFA array.

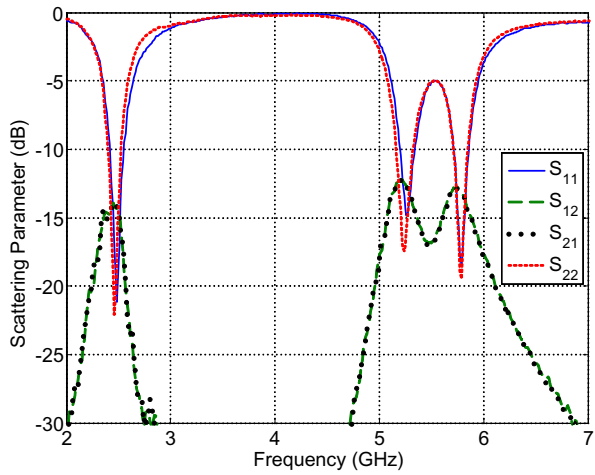

Fig. 6: Measured scattering parameters for the 2-element PIFA array.

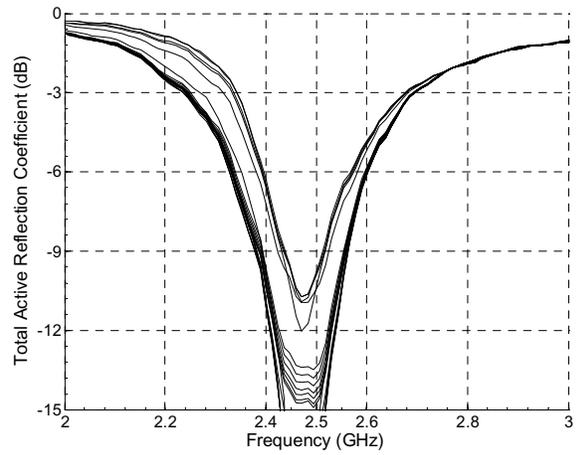

Fig. 7: TARC for one element of the 2-element PIFA array. Each curve represents the measured reflection coefficient for an excitation with constant amplitude but different

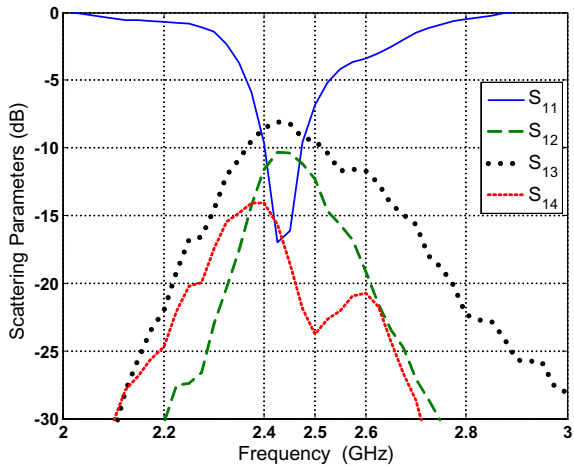

Fig. 8: First row of the measured scattering matrix of the four-element compact array.

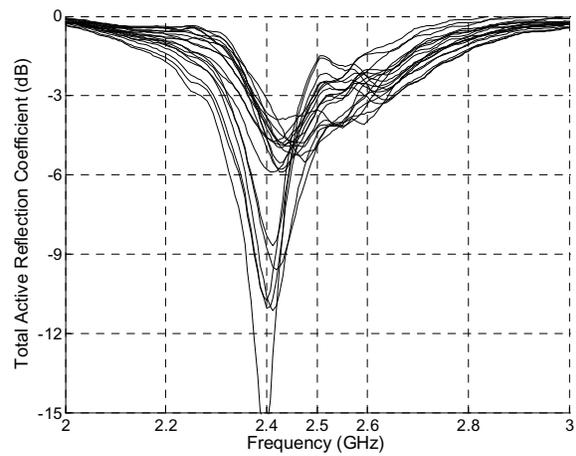

Fig. 9: TARC for one element of the 4-element PIFA array. Each curve represents the measured reflection coefficient for an excitation with constant amplitude but different random phase.

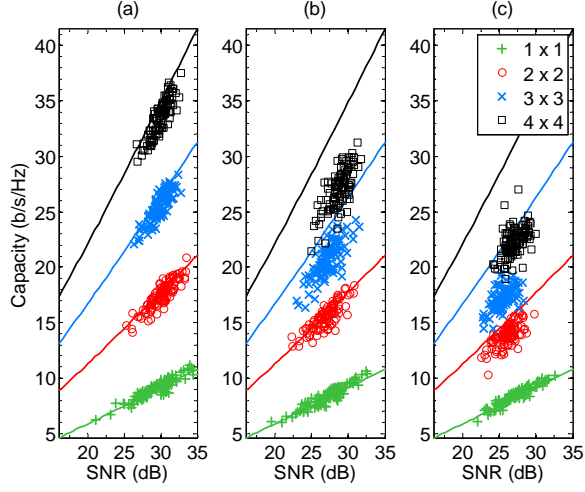

Fig. 10: Measured channel capacity for $\{L \times L \mid L \in [1,2,3,4]\}$ MIMO cases with dipole arrays for which the dipole separations, denoted $(\Delta_d^{(TX)}, \Delta_d^{(RX)})$, were: (a) $(\lambda, \lambda)$, (b) $(\lambda/8, \lambda)$, and (c) $(\lambda/8, \lambda/8)$. The scenario was NLOS in an empty hall. The ergodic capacities for i.i.d. Gaussian channels are plotted as solid curves.

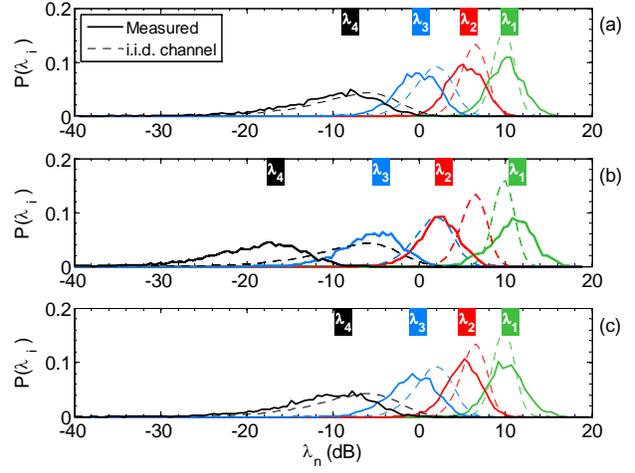

Fig. 11: Distributions of ordered eigenvalues of $\mathcal{HH}^H$ for $4 \times 4$ MIMO. (a) dipole ULA arrays with $\Delta_d^{(TX)} = \lambda$ and $\Delta_d^{(RX)} = \lambda$, (b) dipole ULA arrays with $\Delta_d^{(TX)} = \lambda$ and $\Delta_d^{(RX)} = \lambda/8$, (c) dipole ULA TX array $\Delta_d^{(TX)} = \lambda$ and 4-element PIFA RX array. The scenario was NLOS in an empty hall. The ordered eigenvalue distributions for an i.i.d. Gaussian channel are plotted as dashed curves. Eigenvalue labels are positioned over the mean of the measured eigenvalue curves.

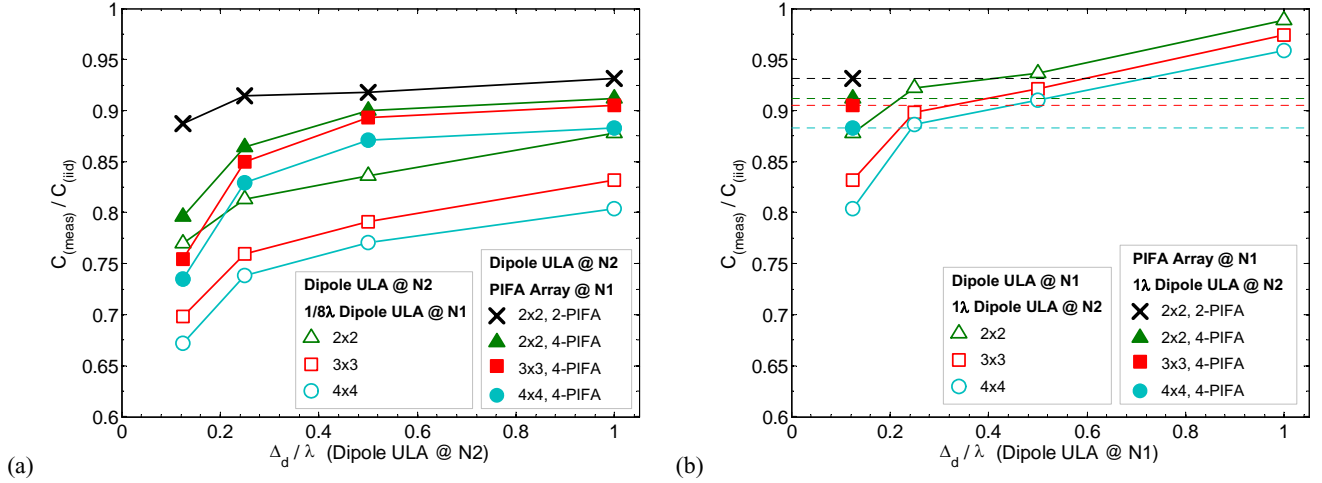

Fig. 12: Normalized MIMO capacity averaged over all measurement sites. (a) Node 1 has fixed arrays ($\lambda/8$ Dipole ULA or compact PIFA Arrays) and Node 2 has a shrinking Dipole ULA. (b) Node 1 has a shrinking Dipole ULA and Node 2 has a fixed $1\lambda$ Dipole ULA. Also shown in (b) are data for when the Dipole array at Node 1 was replaced with compact PIFA Arrays.